\documentclass[prd,a4paper,showpacs,showkeys,preprint,byrevtex]
{revtex4}
\usepackage{graphicx}
\usepackage{dcolumn}
\usepackage{amsmath}
\usepackage{bm}

\begin{document}

\title{Semirelativistic potential model for low-lying three-gluon
glueballs}

\author{Vincent \surname{Mathieu}}
\thanks{IISN Scientific Research Worker}
\email[E-mail: ]{vincent.mathieu@umh.ac.be}
\author{Claude \surname{Semay}}
\thanks{FNRS Research Associate}
\email[E-mail: ]{claude.semay@umh.ac.be}
\affiliation{Groupe de Physique Nucl\'{e}aire Th\'{e}orique,
Universit\'{e} de Mons-Hainaut,
Acad\'{e}mie universitaire Wallonie-Bruxelles,
Place du Parc 20, BE-7000 Mons, Belgium}
\author{Bernard \surname{Silvestre-Brac}}
\email[E-mail: ]{silvestre@lpsc.in2p3.fr}
\affiliation{Laboratoire de Physique Subatomique et de Cosmologie,
Avenue des Martyrs 53, FR-38026 Grenoble-Cedex, France}

\date{\today}

\begin{abstract}
The three-gluon glueball states are studied with the generalization of a
semirelativistic potential model giving good results for two-gluon
glueballs. The Hamiltonian depends only on 3 parameters fixed on
two-gluon glueball spectra: the strong coupling constant, the string
tension, and a gluon size which removes singularities in the potential.
The Casimir scaling determines the structure of the confinement.
Low-lying $J^{PC}$ states are computed and compared with recent lattice
calculations. A good agreement is found for $1^{--}$ and $3^{--}$
states, but our model predicts a $2^{--}$ state much higher in energy
than the lattice result. The $0^{-+}$ mass is also computed.
\end{abstract}


\maketitle

\section{Introduction}
\label{sec:intro}

The QCD theory allows the existence of bound states of gluons, called
glueballs, but no firm experimental discovery of such states has been
obtained yet. An important difficulty is that glueball states might
possibly mix strongly with nearby meson states. Nevertheless, the
computation of pure gluon glueballs remains an interesting task. This
could guide experimental searches and provide some calibration for more
realistic models of glueballs.

Lattice calculations are undoubtedly a powerful tool to investigate the
structure of glueballs. A previous study \cite{morn99} predicts the
existence of a lot of resonances between 2 and 4~GeV. A recent update of
this work \cite{chen06} confirms the results already obtained.

The potential model, which is so successful to describe bound states of
quarks, is also a possible approach to study glueballs
\cite{corn83,brau05,hou84,hou01}. In a recent paper \cite{brau04},
a semirelativistic Hamiltonian is used to compute two-gluon glueballs
with masses in good agreement with those obtained by the lattice
calculations of Ref.~\cite{morn99}. This Hamiltonian, the model III in
Ref.~\cite{brau04}, relies on the
auxiliary fields formalism \cite{morg99,sema04} and on a one-gluon
exchange (OGE) interaction proposed in Ref.~\cite{corn83}. It depends
only on three parameters: the strong coupling constant $\alpha_S$, the
string tension $a$, and a gluon size $\gamma$ which removes
singularities in the short-range
part of the potential. The constituent gluon mass is dynamically
generated and it is assumed that the Casimir scaling determines the
color structure of the confinement. These two ingredients are actually
necessary to obtain a good agreement between the results from a
potential model and from lattice calculations.

The purpose of this paper is to check if the potential model built for
two-gluon systems in Ref.~\cite{brau04} can be generalized to
three-gluon systems. Compared to previous models \cite{hou84,hou01}, our
approach is characterized by some improved features: semirelativistic
kinematics, more realistic confinement, dynamical definition of the
gluon mass, coherent treatment of the gluon size. These points will be
detailed below. The masses of the lowest negative parity $L=0$ glueballs
are computed with a great accuracy and compared with lattice
calculations
\cite{morn99,chen06}. In Sec.~\ref{sec:ham}, the three-gluon Hamiltonian
is build, and the structure of the glueballs studied is presented in
Sec.~\ref{sec:wf}. The three-gluon glueball spectrum is presented with
the two-gluon glueball spectrum from Ref.~\cite{brau04} and is discussed
in Sec.~\ref{sec:res}. Some concluding remarks are given in
Sec.~\ref{sec:conc}.

\section{Hamiltonian}
\label{sec:ham}

\subsection{Parameters}
\label{ssec:param}

In Ref.~\cite{brau04}, two sets of parameters, denoted A and B, were
presented for the model III (see Table~\ref{tab:par}). With the set A,
it is possible to obtain glueball masses in agreement with the results
of some experimental works \cite{zou99,bugg00}: the lowest $2^{++}$
state near 2~GeV, the lowest $0^{++}$ state near 1.5~GeV, and the lowest
$0^{-+}$ state near 2.1~GeV. The values of $a$ and $\alpha_S$ are close
to the ones used in some recent baryon calculations \cite{naro02}. With
the set B, glueball masses were computed in agreement with the results
of the lattice calculations of Ref.~\cite{morn99}. If the
absolute glueball masses found in Ref.~\cite{brau04} with both sets are
strongly different, the relative spectra are nearly identical. As we use
in this work a three-body generalization of the Hamiltonian model III of
Ref.~\cite{brau04}, the two sets will also be considered.

It is worth mentioning how the parameters have been determined in
Ref.~\cite{brau04}. The mass of the lightest $2^{++}$ is nearly
independent of the values of $\alpha_S$ and $\gamma$, but depends
strongly on $a$. So, this last parameter has been determined with this
$2^{++}$ state. The remaining parameters $\alpha_S$ and $\gamma$ have
then be
computed in order to reproduce the lightest $0^{++}$ and $0^{-+}$
states. The three states $2^{++}$, $0^{++}$, and $0^{-+}$ have been
chosen because they are possible experimental glueball candidates
\cite{zou99,bugg00} and because they are computed with relatively small
errors in lattice calculations \cite{morn99,chen06}.

\subsection{Confinement potential}
\label{ssec:conf}

A good approximation of the confining interaction between a quark and an
antiquark in a meson is given by the linear potential $a\, r$, where $r$
is the distance between the two particles and where $a$ is the string
tension. In a baryon, lattice calculations and some theoretical
considerations indicate that each quark generates a flux tube and that
these flux tubes meet in a junction point $\bm R_0$ which minimizes the
potential energy. Following this hypothesis, the confinement in a baryon
could be simulated by the three-body interaction
\begin{equation}
\label{vyqqq}
V_{qqq} = a \sum_{i=1}^{3} |\bm r_i- \bm R_0|,
\end{equation}
For such a potential, the point $\bm R_0$ minimizes also the
length of the three flux tubes and is identified with the Toricelli
point \cite{silv04}.

The energy density $\lambda_c$ of a flux tube (string tension) can
depend on the color
charge $c$ which generates it. Lattice calculations \cite{deld00} and
effective models of QCD \cite{sema04a} predict that the Casimir scaling
hypothesis is well verified in QCD, that is to say that the energy
density is proportional to the value of the quadratic Casimir operator
$\hat F_c^2$ of the colour source
\begin{equation}
\label{la}
\lambda_c = \hat F_c^2 \, \sigma.
\end{equation}
We have then $\lambda_q = \lambda_{\bar q} = 4\,\sigma/3 = a$ and
$\lambda_g = 3\,\sigma$.
In this work we will assume that the confinement
in a three-body color singlet is given by
\begin{equation}
\label{vyccc}
V_{ccc} = \sigma \sum_{i=1}^{3} \hat F_i^2\,|\bm r_i- \bm R_0|.
\end{equation}
This potential can be considered as the three-body generalization of the
confinement used in Ref.~\cite{brau04}. No constant potential is added,
contrary to usual Hamiltonians in mesons and baryons
\cite{corn83,simo01}.
Let us note that if the three color charges are not the same, $\bm R_0$
is no longer identified with the Toricelli point \cite{math06}.

Interaction~(\ref{vyccc}) is very difficult to use in a practical
calculation. A good approximation can be obtained for three identical
color charges by replacing $\bm R_0$ by the center of mass coordinate
$\bm R_{\text{cm}}$ and by renormalizing the potential by a factor $f$
which depends on the three-body system \cite{silv04}. For three
identical particles, the best value is $f=0.9515$. We will use this
approximation in the following, which seems more realistic than a
confinement obtained by the sum of two-body forces \cite{hou84,hou01}.

As already mentioned, only the $L=0$ states are studied in this paper.
So, no spin-orbit correction to the confinement \cite{brau04} is taken
into account here.

In Refs.~\cite{corn83,hou01,hou84}, the confinement potential
saturates
at large distances in order to simulate the breaking of the color flux
tube between gluons due to color screening effects. An interaction of
type~(\ref{vyccc}) seems a priori inappropriate since the potential
energy can grow without limit. But the phenomenon of flux tube breaking
must only contribute
to the masses of the highest glueball states. Moreover, it has been
shown that the introduction of a saturation could not be the best
procedure to simulate the breaking of a string joining two colored
objects \cite{swan05}.

\subsection{Dynamical constituent gluon mass}
\label{ssec:dyn}

Within the auxiliary field formalism (also called einbein field
formalism) \cite{morg99}, which can be considered as an approximate way
to handle semirelativistic Hamiltonians \cite{sema04,buis04}, the
effective QCD
Hamiltonian has a kinetic part depending on the current particle masses
$m_i$ and the interaction is dominated by the confinement. A state
dependent constituent mass
$\mu_i = \langle \sqrt{\bm p_i^2+m_i^2} \rangle$ can be defined for each
particle, and all relativistic corrections (spin, momentum, \ldots) to
the static potentials are then expanded in powers of $1/\mu_i$. This
approach has been used in Ref.~\cite{brau04} to build the two-gluon
Hamiltonian. So, the same formalism will be applied also in this paper.

Taking into account the considerations of Sec.~\ref{ssec:conf}, the
simplest generalization to a three-gluon system of the dominant part of
the model III two-gluon Hamiltonian of Ref.~\cite{brau04} is
\begin{equation}
\label{H0_3g}
H_0 = \sum_{i=1}^{3}\sqrt{\bm p^2_i} + f\, \sigma \sum_{i=1}^3
\hat F_i^2\,|\bm r_i- \bm R_{\text{cm}}|,
\end{equation}
with the condition $\sum_{i=1}^{3}\bm p_i = \bm 0$, since we work in the
center of mass of the glueball. The gluons have vanishing current
masses and their color is such that $\langle \hat F_i^2 \rangle=3$.
Contrary to some previous works \cite{corn83,hou01,hou84}, our
Hamiltonian is a
semirelativistic one. In Ref.~\cite{brau04}, it has been shown that it
is an important ingredient to obtain correct two-gluon glueball spectra.

Using the technics of Ref.~\cite{kerb00}, it is
possible to obtain an analytical approximate formula giving the glueball
mass $M_0$ and the
constituent gluon mass $\mu_0$ (the three constituent gluon masses are
the same
since the wave function is completely symmetrized,
see Sec.~\ref{sec:wf}) for the $L=0$ eigenstates of Hamiltonian $H_0$
\begin{equation}
\label{mu_3g}
M_0 \approx 6\, \mu_0 \quad \text{with} \quad
\mu_0 \approx 2\sqrt{\frac{6\, f\, \sigma}{\pi}} \left[
\frac{2}{3\times 5^{1/3}} + \frac{8}{9\times 5^{5/6}}
\left( n + \frac{1}{2} \right) \right]^{3/4}.
\end{equation}
The accuracy of these formulas, which is around 5\% for the ground state
($n=0$), deteriorates with increasing values of $n$. With a value of
$a=4\, \sigma/3$ around 0.2~GeV$^2$, the smallest gluon constituent
mass is around 600~MeV. It is then relevant to use an expansion in
powers of $1/\mu_0$. Such a value of the gluon mass is in
agreement with the values used in Refs.~\cite{corn83,hou01,hou84},
but here the constituent mass is dynamically generated.

Instead of using the auxiliary field formalism, it is possible to
consider relativistic corrections which are expanded in powers of
$1/E_i(\bm p_i)$ where $E_i(\bm p_i)=\sqrt{\bm p_i^2+m_i^2}$ (see for
instance Ref.~\cite{godf85}). But, this leads to very complicated
non local potentials which are difficult to handle.

\subsection{Short-range potential}
\label{ssec:sr}

The Hamiltonian $H_0$~(\ref{H0_3g}) gives the main features of the
three-gluon glueball spectra, but the introduction of a short-range
potential is necessary to achieve a detailed study. In
Ref.~\cite{brau04}, a OGE interaction between two gluons, coming from
Ref.~\cite{corn83}, has been considered. It is not possible to use it
directly for a three-gluon glueball because the color structure of the
interaction is different. So, we use here the last version of a OGE
interaction between two gluons developed specifically for three-gluon
glueballs \cite{hou84,hou01}. Its explicit form, which is very similar
to the form of the OGE interaction for two-gluon glueballs, is given
below.

This interaction contains a tensor part and a spin-orbit part. Both are
neglected in this paper since only $L=0$ states are studied. Moreover,
it has been shown that the tensor interaction between two gluons is
small in two-gluon glueballs \cite{brau04}.

The OGE two-gluon potential has a priori a very serious flaw: depending
on the spin state, the short-range singular part of the potential may be
attractive and lead to a Hamiltonian unbounded from below \cite{hou01}.
This problem is solved, as in Ref.~\cite{brau04}, by giving a finite
size to the gluon (see Sec.~\ref{ssec:size}).

The OGE two-gluon potential depends on the gluon constituent mass. To
determine it, we follow the procedure proposed in Ref.~\cite{brau04}.
For a given set of quantum numbers $\{\alpha\}$, the eigenstate
$|\phi_\alpha\rangle$ of the Hamiltonian $H_0$ is computed. With this
state, a constituent gluon mass is computed
$\mu_\alpha = \langle \phi_\alpha |\sqrt{\bm p_1^2}|\phi_\alpha
\rangle$. This value of $\mu_\alpha$ is then used in the complete
Hamiltonian (see Sec.~\ref{ssec:ht}) to compute its eigenstate with
quantum numbers
$\{\alpha\}$. It is worth noting that, with this procedure, two states
which differ only by the radial quantum number are not orthogonal
since they are eigenstates of two different Hamiltonians which differ by
the value of $\mu$. It is shown in Ref.~\cite{sema04} that this problem
is not serious, the overlap of these states being generally weak.

\subsection{Gluon size}
\label{ssec:size}

In potential models, the gluon is considered as an effective degree of
freedom with a constituent mass. Within this framework, it is natural to
assume that a gluon is not a pure pointlike particle but an object
dressed by a gluon and quark-antiquark pair cloud. Such an
hypothesis for quarks leads to very good results in meson
\cite{brau98} and baryon \cite{brau02} sectors. As in
Ref.~\cite{brau04}, we assume here a Yukawa color charge density for the
gluon
\begin{equation}
\rho(\bm u)=\frac{1}{4\pi \gamma^2}
\frac{e^{-u/\gamma}}{u},
\label{dens}
\end{equation}
where $\gamma$ is the gluon size parameter.
The interactions between gluons are then modified by this density, a
bare potential being transformed into a dressed one.

The main purpose of the gluon dressing is to remove all singularities
in the short-range part of the interaction \cite{brau02}.
But, for consistency, the same regularization is applied to the
confinement potential, although no singularity is present in this case.
We think that the definition of a gluon size, which has a clear physical
meaning, is preferable to the use of
a smearing function only for potentials with singularity
\cite{brau05,hou01}.

A one-body potential, like $|\bm r_i-\bm R_{\text{cm}}|$, is dressed
by a simple convolution over the density of
the interacting gluon and the potential
\begin{equation}
\label{conv1}
V(\bm r)^*=\int d\bm r'\,V(\bm r')\, \rho(\bm r-\bm r').
\end{equation}
A dressed two-body potential, depending on $|\bm r_i-\bm r_j|$, is
obtained by a double convolution. This procedure is equivalent to the
following calculation
\cite{sema03}
\begin{equation}
\label{conv2}
V(\bm r)^{**}=\int d\bm r'\,V(\bm r')\, \Gamma(\bm r-\bm r')
\quad \text{with} \quad
\Gamma(\bm u)=\frac{1}{8\pi \gamma^3}e^{-u/\gamma}.
\end{equation}

\subsection{Total Hamiltonian}
\label{ssec:ht}

To obtain the total Hamiltonian for three-gluon glueballs which is the
simplest generalization of the Hamiltonian for two-gluon glueballs from
Ref.~\cite{brau04}, we take the
Hamiltonian $H_0$ given by the relation~(\ref{H0_3g}); we add the OGE
interactions coming from Ref.~\cite{hou01} (without spin-orbit and
tensor parts); and we dress all the potentials with the gluon color
density~(\ref{dens}). This gives the following Hamiltonian
\begin{subequations}
\label{h_3g}
\begin{eqnarray}
H &=& \sum_{i=1}^{3}\sqrt{\bm p^2_i}+ V_{\text{OGE}}^{**}+
V_{\text{Conf}}^* \quad \text{with} \label{h_3ga}\\
V_{\text{OGE}}^{**} &=&\alpha_S \sum_{i<j=1}^3 \hat F_i\cdot\hat
F_j \left[\left(\frac{1}{4}+\frac{1}{3}\vec S\,^2_{ij}\right)
U(r_{ij})^{**} -\frac{\pi}{\mu^2}\delta(\bm r_{ij})^{**}
\left(\beta+\frac{5}{6}\vec S\,^2_{ij}\right)\right],\label{h_3gb}\\
U(r)^{**}&=&\frac{1}{(\mu^2\gamma^2-1)^2}\left(\frac{e^{-\mu r}}{r}
- \frac{e^{-r/\gamma}}{r}\right) +
\frac{e^{-r/\gamma}}{2\gamma(\mu^2\gamma^2-1)}
\quad \text{with} \quad U(r)=\frac{e^{-\mu r}}{r},\label{h_3gc}\\
\delta(\bm r)^{**}&=&\frac{1}{8\pi\gamma^3}e^{-r/\gamma},\label{h_3gd}
\\
V_{\text{Conf}}^*&=&f\,\sigma\sum_{i=1}^3 \hat F_i^2\,
|\bm r_i- \bm R_{\text{cm}}|^* \quad \text{with} \quad
r^*=r+2\gamma^2\frac{1-e^{-r/\gamma}}{r},\label{h_3ge}
\end{eqnarray}
\end{subequations}
where $\sigma=3\, a/4$, $\sum_{i=1}^{3}\bm p_i = \bm 0$ and
$\vec S_{ij}=\vec S_i+\vec S_j$.
$\beta=+1$ ($-1$) for a gluon pair in color octet antisymmetrical
(symmetrical) state. The constituent state-dependent gluon mass $\mu$ is
computed in advance with a solution of the Hamiltonian $H_0$.

\section{Wave functions}
\label{sec:wf}

A gluon is a $I(J^P)=0(1^-)$ color octet state.
Two different three-gluon color singlet states exist \cite{hou84}, which
are completely symmetrical or completely antisymmetrical
($\langle \hat F_i\cdot\hat F_j\rangle = 3$ for such states). The total
isospin state of a glueball is an isosinglet and is completely
symmetrical. Different total spin states are allowed with different
properties of symmetry. They are presented in Table~\ref{tab:s}. As
gluons are bosons, the total wave function must be completely
symmetrical. Its parity is the opposite of the spatial parity, and its
$C$-parity is positive for
color antisymmetrical state and negative for color symmetrical state.
Let us note that a two-gluon glueball has always a positive $C$-parity.

In this work, we will mainly consider glueballs with the lowest masses.
These states are characterized by a vanishing total orbital angular
momentum $L=0$ and by a spatial wave function completely symmetrical
with a positive parity. This immediately implies that the lowest
glueballs are states with $J^{PC}$ equal to $0^{-+}$, $1^{--}$, and
$3^{--}$ \cite{hou84,hou01}. In order to reach a good accuracy, the
trial spatial wave functions are expanded in large gaussian function
bases \cite{suzu98}. With more than 10 gaussian functions for each
color/isospin/spin channels, we have checked that the numerical errors
on masses presented are around or less than 1~MeV.

Using the value of $\sigma$ from models A and B, the $L=0$ ground state
masses of the Hamiltonian $H_0$~(\ref{H0_3g}) are presented in
Table~\ref{tab:mu} for all possible $J^{PC}$ quantum numbers. The
$0^{-+}$, $1^{--}$, and $3^{--}$ glueballs have clearly the lowest
masses. In this table, they are degenerate since the Hamiltonian $H_0$
is spin-independent. For each state, the corresponding constituent gluon
masses $\mu_0$ is indicated. It is used to define the complete
Hamiltonian $H$~(\ref{h_3g}).

\section{Results}
\label{sec:res}

\begin{center}
\begin{figure}
\includegraphics*[height=8cm]{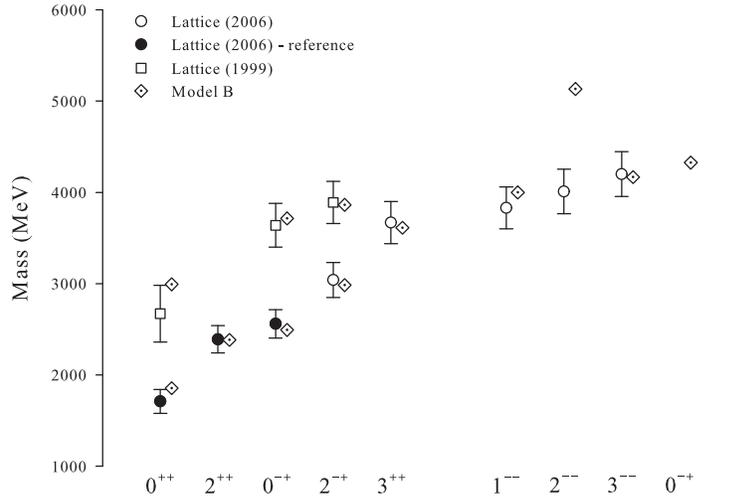}
\protect\caption{Glueball masses given in MeV. Dotted diamonds: Results
from model B (two-gluon masses are taken from
Ref.~\cite{brau04}); Black and white circles: Lattice results from
Ref.~\cite{morn99}; White squares: Lattice results from
Ref.~\cite{chen06}. Black circles indicate the reference states taken as
inputs to fix the parameters. The error bars for lattice results are
computed by summing the two uncertainties (see Table~\ref{tab:m}).
}
\label{fig1}
\end{figure}
\end{center}

We present here the three-gluon glueball masses obtained with the
complete Hamiltonian $H$~(\ref{h_3g}) together with the two-gluon
glueball masses
computed in Ref.~\cite{brau04} (see Table~\ref{tab:m}). These masses are
compared with the results obtained by the lattice calculations of
Ref.~\cite{chen06}. This work is an update of a previous study
\cite{morn99}. So, a state not computed in Ref.~\cite{chen06} but
presented in Ref.~\cite{morn99} is also considered here. As it can be
seen on Fig.~\ref{fig1}, with the set B
of parameters, our masses are in quite good agreement with the results
of the lattice calculations, except for one exception discussed below.
Unfortunately, the masses predicted for the $0^{++}$, $2^{++}$, and
$0^{+-}$ two-gluon glueballs are larger than some possible experimental
candidates \cite{zou99,bugg00}. A best agreement with these data can be
achieved with the set A of parameters \cite{brau04}. But the situation
is not simple, since recent works \cite{bicu06,bugg06} suggest that the
new observed resonance $f_0(1810)$ reported by the BES collaboration
\cite{abli06} could be a $0^{++}$ glueball. Such a
data is then compatible with the set B of parameters.

\begin{center}
\begin{figure}
\includegraphics*[height=8cm]{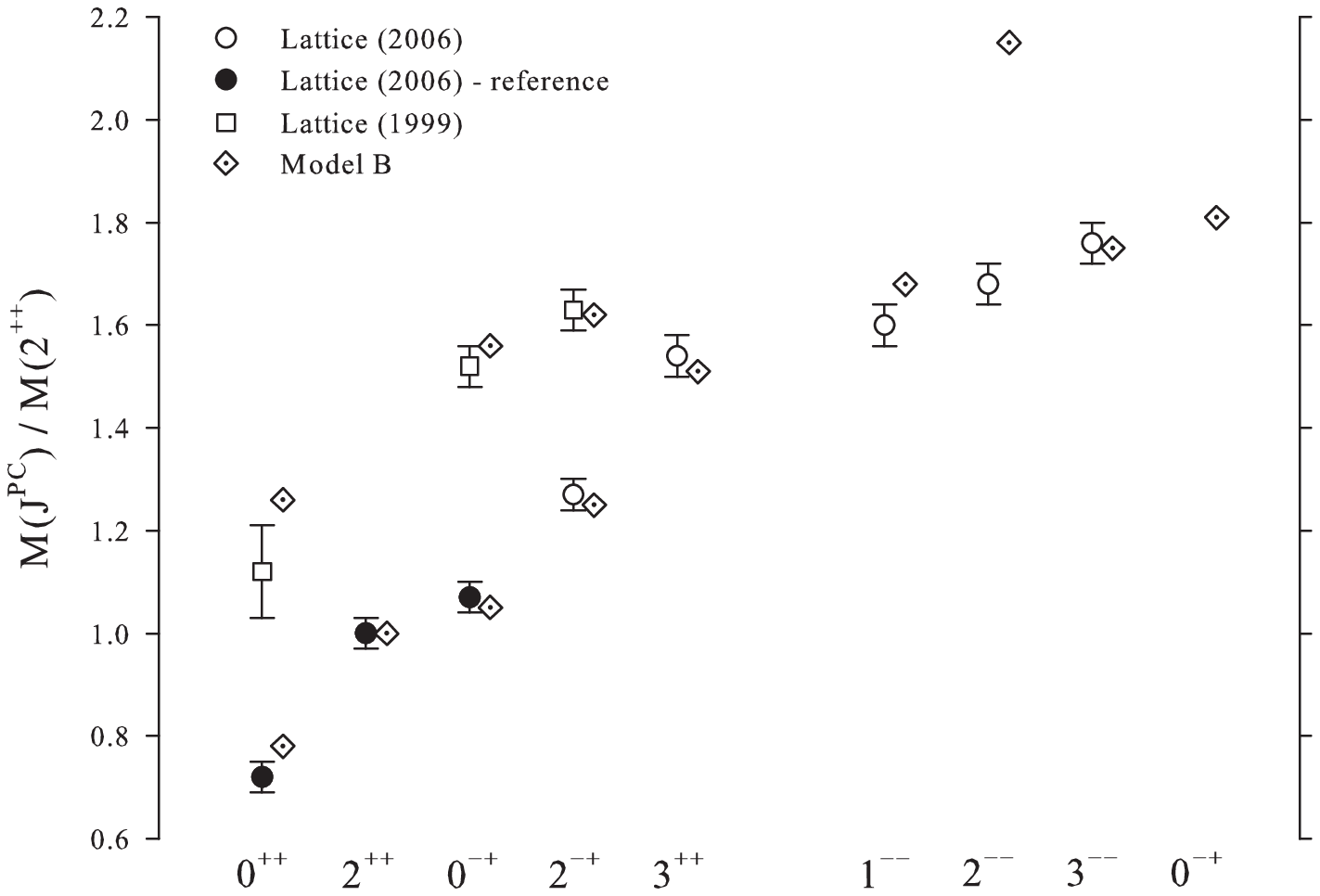}
\protect\caption{Glueball mass ratios normalized to the lightest
$2^{++}$ state (see Table~\ref{tab:m}). Dotted diamonds: Results from
model B (two-gluon masses are taken from
Ref.~\cite{brau04}); Black and white circles: Lattice results from
Ref.~\cite{morn99}; White squares: Lattice results from
Ref.~\cite{chen06}. Black circles indicate the reference states taken as
inputs to fix the parameters. The error bars for lattice results are
computed without the normalization error on glueball masses. Results
from models A and B cannot be distinguished on this graphic.
}
\label{fig2}
\end{figure}
\end{center}

As the lattice results can suffer from quite large error of
normalization, it is then interesting to present a relative spectra. In
Fig.~\ref{fig2}, the ratios of two- and three-gluon glueball masses on
the lowest $2^{++}$ mass are presented. At this scale, results coming
from sets A and B of parameters cannot be distinguished.

Let us first point that the $1^{--}$ and $3^{--}$ states occur only in
three-gluon glueballs, whereas the $0^{-+}$ state is available for two-
and three-gluon systems. The mixing between these two channels is
ignored here. Our results share some similarities with other potential
models. For instance, in Ref.~\cite{hou84}, the three lowest three-gluon
glueballs are those with $J^{PC}$ equal to $0^{-+}$, $1^{--}$, and
$3^{--}$, but they are found around 2400~MeV and the mass splitting
between theses states is around 50~MeV. In Ref.~\cite{hou01}, the same
three lowest states are found: the $1^{--}$ glueball is predicted in the
range 3500-3700~MeV and the mass difference between these three sates is
around 100~MeV.

A detailed comparison between the results of our potential model and the
lattice calculations of Ref.~\cite{morn99} have been performed in
Ref.~\cite{brau04} for the
two-gluon glueballs. The conclusions of this work are not changed by
the new lattice predictions obtained in Ref.~\cite{chen06}. Let us then
focus our attention on the three-gluon glueball spectrum obtained
with the set B of parameters. Without any new
parameters, the $1^{--}$ and $3^{--}$ glueballs are in good agreement
with the lattice predictions of Ref.~\cite{chen06} (see
Table~\ref{tab:m} and Fig.~\ref{fig1}). The agreement is slightly less
good for the relative spectra (see Fig.~\ref{fig2}) because the error is
smaller for lattice mass ratios. Our results also suggest
the existence of a three-gluon $0^{-+}$ glueball near these two last
states. The others $0^{-+}$ states already computed by the lattice
calculations can be identified as two-gluon systems by our model.

The lattice results predict a $2^{--}$ state at 4010~MeV near the
$1^{--}$ and $3^{--}$ states. With our Hamiltonian, a mass more than
1~GeV above is computed. It is unavoidable in our model, since a spin 2
function has a mixed symmetry which implies a mixed symmetry for the
space function and then a greater mass for the corresponding glueball,
in agreement with the results of Refs.~\cite{hou84,hou01}. It has been
checked that the spatial wave function of the $2^{--}$ state is
dominated by a configuration in which each internal variable is
characterized by one unit of angular momentum.

We have no
firm explanation for such a discrepancy. This problem could arise
because the gluon has a constituent mass within our formalism. So, it
possesses a spin as any massive particle, that is to say three states of
polarization. In lattice calculations, the gluon is a massless particle
with a definite helicity and then only two states of polarization. The
same phenomenon could be at the origin of another puzzling--to some
extent opposite--situation in the two-gluon glueball sector: the
existence of $1^{PC}$ states with low masses in our model \cite{brau04},
which are actually not predicted below 4~GeV by lattice calculations
\cite{morn99}. Further studies are necessary to clarify the situation.

The lattice results predict also several three-gluon $J^{+-}$ states in
the range 2980-4780~MeV. These states with positive parity have a
negative spatial parity and then are expected to have masses larger than
the lowest three-gluon $J^{--}$ states. This is manifestly not the case
in these lattice calculations. In Ref.~\cite{brau04}, it is
shown that the structure of some two-gluon glueballs can be explained
with our potential model by the existence of strong spin-orbit forces
coming from the OGE interaction and the confinement. It is possible that
similar forces act in three-gluon system to lower the masses of some of
these states.

\section{Conclusion}
\label{sec:conc}

The masses of pure three-gluon glueballs have been studied with the
generalization of a semirelativistic potential model \cite{brau04} which
gives pure two-gluon glueball spectra in good agreement with
lattice calculations \cite{morn99,chen06}. The short-range part of the
potential is the sum of two-body OGE interactions. For the confinement,
a potential simulating a genuine Y-junction is used and it is assumed
that the Casimir scaling hypothesis is well verified. The gluon is
massless but the OGE interaction is expressed in terms of a state
dependent constituent mass. The Hamiltonian depends only on 3~parameters
fixed in Ref.~\cite{brau04}: the strong coupling constant, the string
tension, and a gluon size. All masses have been accurately computed with
an expansion of trial states in gaussian functions \cite{suzu98}.

In this work, only the negative (natural) parity $L=0$ three-gluon
glueballs are studied. The masses of the lowest $1^{--}$ and $3^{--}$
glueballs predicted by our potential model are in agreement with the
results of a recent lattice calculations \cite{chen06}, but the lowest
$2^{--}$ state is found at higher energy in agreement with other
potential models \cite{hou84,hou01}. The origin of such a discrepancy
between both approaches is not known. It could be due to the fact that
gluons have constituent non vanishing masses
in our approach. They are then characterized by a spin, and not by a
helicity as it could be expected for particles with a vanishing current
mass.

Other positive parity three-gluon glueballs predicted by lattice
calculations are not considered here \cite{chen06}. We think that their
properties could be explained by the action of strong spin-orbit forces,
similar to those present into two-gluon glueballs. To take into account
these interaction, it is necessary to consider the spin-orbit forces
arising from OGE interactions but also from the Y-junction confinement.
Such a work is in progress.

\section*{acknowledgments}

The authors would thank F.~Buisseret for useful discussions.
V.~Mathieu (IISN Scientific Research Worker) and C.~Semay (FNRS Research
Associate) would like to thank the FNRS for financial support.


\clearpage

\begin{table}[h]
\protect\caption{Parameters for models A and B ($\sigma=3\,a/4$). For
both models, the gluon current mass is zero and $f=0.9515$.}
\label{tab:par}
\begin{ruledtabular}
\begin{tabular}{lll}
& Model A & Model B \\
\hline
$a$ & 0.16~GeV$^2$ & 0.21~GeV$^2$ \\
$\alpha_S$ & 0.40 & 0.50 \\
$\gamma$ & 0.504~GeV$^{-1}$ & 0.495~GeV$^{-1}$ \\
\end{tabular}
\end{ruledtabular}
\end{table}

\begin{table}[h]
\protect\caption{Characteristics of three-gluon spin functions with
total spin $S$, intermediate couplings $S_{\text{int}}$, and symmetry
properties which can be obtained by coupling
(A: Antisymmetrical, S: Symmetrical, MS: Mixed symmetry).}
\label{tab:s}
\begin{ruledtabular}
\begin{tabular}{lll}
$S$ & $S_{\text{int}}$ & Symmetry \\
\hline
0 & 0 & 1 A \\
1 & 0, 1, 2 & 1 S, 2 MS \\
2 & 1, 2 & 2 MS \\
3 & 2 & 1 S
\end{tabular}
\end{ruledtabular}
\end{table}

\begin{table}[h]
\protect\caption{$L=0$ ground state masses $M_0$ of the
Hamiltonian $H_0$~(\ref{H0_3g}) as a function of the $J^{PC}$ quantum
numbers.
The corresponding constituent gluon masses $\mu_0$ are also
given. Values in MeV are computed with the value of $\sigma$ from models
A/B. The lowest masses are printed in italic.}
\label{tab:mu}
\begin{ruledtabular}
\begin{tabular}{llllll}
$J^{PC}$ & $M_0$ & $\mu_0$ & $J^{PC}$ & $M_0$ & $\mu_0$ \\
\hline
$0^{--}$ & 5574/6385 & 929/1064 & $0^{-+}$ & \emph{3211}/\emph{3679} &
535/613 \\
$1^{--}$ & \emph{3211}/\emph{3679} & 535/613 & $1^{-+}$ & 4156/4761 &
693/794
\\
$2^{--}$ & 4156/4761 & 693/794 & $2^{-+}$ & 4156/4761 & 693/794 \\
$3^{--}$ & \emph{3211}/\emph{3679} & 535/613 & $3^{-+}$ & 5574/6385 &
929/1064 \\
\end{tabular}
\end{ruledtabular}
\end{table}

\begin{table}[h]
\protect\caption{Glueball masses in MeV and (glueball mass ratios
normalized to lightest $2^{++}$). The two-gluon masses are taken from
Ref.~\cite{brau04}. The error bars for lattice mass ratios
are computed without the normalization error on the masses. The lightest
$0^{++}$, $2^{++}$, and $0^{-+}$ states are taken as inputs to fix the
parameters. The first column indicates the valence gluon content as
predicted by our model.}
\label{tab:m}
\begin{ruledtabular}
\begin{tabular}{lllll}
 &$J^{PC}$ & Lattice [Ref.] & Model A & Model B \\
\hline
gg&$0^{++}$ & 1710$\pm$50$\pm$80\phantom{00} ($0.72\pm0.03$)
\cite{chen06} & 1604 (0.78) & 1855 (0.78) \\
&& 2670$\pm$180$\pm$130 ($1.12\pm0.09$) \cite{morn99} & 2592 (1.26)
& 2992 (1.26) \\
&$2^{++}$ & 2390$\pm$30$\pm$120\phantom{0} ($1.00\pm0.03$)
\cite{chen06} & 2051 (1.00) & 2384 (1.00) \\
&$0^{-+}$ & 2560$\pm$35$\pm$120\phantom{0} ($1.07\pm0.03$)
 \cite{chen06} & 2172 (1.06) & 2492 (1.05)  \\
&         & 3640$\pm$60$\pm$180\phantom{0} ($1.52\pm0.04$)
 \cite{morn99} & 3228 (1.57) & 3714 (1.56)\\
&$2^{-+}$ & 3040$\pm$40$\pm$150\phantom{0} ($1.27\pm0.03$)
 \cite{chen06} & 2573 (1.25) & 2984 (1.25) \\
&         & 3890$\pm$40$\pm$190\phantom{0} ($1.63\pm0.04$)
 \cite{morn99} & 3345 (1.63) & 3862 (1.62) \\
&$3^{++}$ & 3670$\pm$50$\pm$180\phantom{0} ($1.54\pm0.04$)
 \cite{chen06} & 3132 (1.53) & 3611 (1.51) \\
 & & & & \\
ggg&$1^{--}$ & 3830$\pm$40$\pm$190\phantom{0} ($1.60\pm0.04$)
 \cite{chen06} & 3433 (1.67) & 3999 (1.68) \\
&$2^{--}$ & 4010$\pm$45$\pm$200\phantom{0} ($1.68\pm0.04$)
 \cite{chen06}& 4422 (2.16) & 5133 (2.15) \\
&$3^{--}$ & 4200$\pm$45$\pm$200\phantom{0} ($1.76\pm0.04$)
 \cite{chen06} & 3569 (1.74) & 4167 (1.75) \\
&$0^{-+}$ &  & 3688 (1.80) & 4325 (1.81)  \\
\end{tabular}
\end{ruledtabular}
\end{table}

\clearpage

\begin{thebibliography}{aa}

\bibitem{morn99} C. J. Morningstar and M. J. Peardon, Phys. Rev. D
{\bf 60}, 034509 (1999) [hep-lat/9901004]. 
\bibitem{chen06} Y.~Chen \emph{et al.}, Phys. Rev. D
{\bf 73}, 014516 (2006) [hep-lat/0510074]. 
\bibitem{corn83} J. M. Cornwall and A. Soni, Phys. Lett. B {\bf 120},
431 (1983).
\bibitem{brau05} F. Brau and C. Semay, Phys. Rev. D \textbf{72}, 078501
(2005) [hep-ph/0411108]. 
\bibitem{hou84} W. S. Hou and A. Soni, Phys. Rev. D {\bf 29}, 101
(1984). 
\bibitem{hou01} W. S. Hou, C. S. Luo, and G. G. Wong, Phys. Rev. D
{\bf 64}, 014028 (2001) [hep-ph/0101146]. 
\bibitem{brau04} F. Brau and C. Semay, Phys. Rev. D \textbf{70}, 014017
(2004) [hep-ph/0412173]. 
\bibitem{morg99} V. L. Morgunov, A. V. Nefediev, and Yu. A. Simonov,
Phys. Lett. B {\bf 459}, 653 (1999) [hep-ph/9906318]. 
\bibitem{sema04} C. Semay, B. Silvestre-Brac, and I. M. Narodetskii,
Phys. Rev. D {\bf 69}, 014003 (2004) [hep-ph/0309256]. 
\bibitem{zou99} B. S. Zou, Nucl. Phys. {\bf A655}, 41 (1999). 
\bibitem{bugg00} D. V. Bugg, M. J. Peardon, and B. S. Zou, Phys. Lett. B
{\bf 486}, 49 (2000) [hep-ph/0006179]. 
\bibitem{naro02} I. M. Narodetskii and M. A. Trusov,
Phys. Atom. Nucl. {\bf 65}, 917 (2002), Yad. Fiz. {\bf 65}, 949 (2002)
[hep-ph/0104019];
Phys. Atom. Nucl. {\bf 67}, 762 (2004), Yad. Fiz. {\bf 67}, 783 (2004)
[hep-ph/0307131].
\bibitem{silv04} B. Silvestre-Brac, C. Semay, I. M. Narodetskii, and A.
I. Veselov, Eur. Phys. J. C \textbf{32}, 385 (2004) [hep-ph/0309247].
\bibitem{deld00} S.~Deldar, Phys. Rev. D {\bf 62}, 034509 (2000)
[hep-lat/9911008]. G.~S.~Bali, Phys. Rev. D {\bf 62}, 114503 (2000)
[hep-lat/0006022]. 
\bibitem{sema04a} C.~Semay, Eur. Phys. J. A {\bf 22}, 353 (2004)
[hep-ph/0409105]. 
T.~H.~Hanson, Phys. Lett. B {\bf 166}, 343 (1986).
\bibitem{simo01} Yu. A. Simonov, Phys. Lett. {\bf 515}, 137 (2001)
[hep-ph/0105141].
\bibitem{math06} V. Mathieu, C. Semay, and F. Brau, Eur. Phys. J. A
{\bf 27}, 225 (2006) [hep-ph/0511210]. 
\bibitem{swan05} E.~S.~Swanson, J. Phys. G: Nucl. Part. Phys. {\bf 31},
845 (2005) [hep-ph/0504097].
\bibitem{buis04} F. Buisseret and C. Semay,
Phys. Rev. D {\bf 70}, 077501 (2004) [hep-ph/0406216]. 
\bibitem{kerb00} B. O. Kerbikov and Yu. A. Simonov, Phys. Rev. D
{\bf 62}, 093016 (2000) [hep-ph/0001243]. 
\bibitem{godf85} S. Godfrey and N. Isgur, Phys. Rev. D {\bf 32}, 189
(1985).
\bibitem{brau98} F. Brau and C. Semay, Phys. Rev. D {\bf 58}, 034015
(1998) [hep-ph/0412179]. 
\bibitem{brau02} F. Brau, C. Semay, and B. Silvestre-Brac,
Phys. Rev. C {\bf 66}, 055202 (2002) [hep-ph/0412176]. 
\bibitem{sema03} B. Silvestre-Brac, F. Brau, and C. Semay,
J. Phys. G: Nucl. Part. Phys. {\bf 29}, 2685 (2003) [hep-ph/0302252].
\bibitem{suzu98} Y. Suzuki and K. Varga, {\em Stochastic variational
approach to quantum mechanical few-body problems} (Springer Verlag,
Berlin, Heidelberg, 1998).
\bibitem{bicu06} P. Bicudo, S. R. Cotanch, F. J. Llanes-Estrada, and D.
Robertson, hep-ph/0602172.
\bibitem{bugg06} D. V. Bugg, hep-ph/0603018.
\bibitem{abli06} M. Ablikim \emph{et. al} (BES Collaboration),
Phys. Rev. Lett. {\bf 96}, 162002 (2006) [hep-ex/0602031].

\end{thebibliography}
\end{document}